\documentclass[12pt]{iopart}

\usepackage{graphicx}

\newcommand{\DL}{\ensuremath{\Delta_{\mathrm{L}}}}
\newcommand{\Ogm}{\ensuremath{0_{\mathrm{g}}^-}}
\newcommand{\Pth}{\ensuremath{\mathrm{P}_{3/2}}}
\newcommand{\Soh}{\ensuremath{\mathrm{S}_{1/2}}}
\newcommand{\Dth}{\ensuremath{\mathrm{D}_{3/2}}}
\newcommand{\aSup}{\ensuremath{\mathrm{a}^3\Sigma_{\mathrm{u}}^{+}}}

\begin{document}

\title[Ultracold cesium molecules by photoassociation with
tunneling]{Efficient formation of strongly bound ultracold cesium
  molecules by photoassociation with tunneling}

\author{Mihaela Vatasescu$^1$, Claude M Dion$^2$ and  Olivier
  Dulieu$^3$}

\address{$^1$ Institute of Space Sciences, MG-23, RO-77125,
  Magurele-Bucharest, Romania} 

\address{$^2$ Department of Physics, Ume{\aa} University, SE-901 87
  Ume{\aa}, Sweden} 

\address{$^3$ Laboratoire~Aim\'{e}~Cotton, CNRS, B\^{a}timent~505,
  Univ. Paris-Sud, 91405~Orsay-Cedex, France}
\ead{olivier.dulieu@lac.u-psud.fr}

\date{\today}

\begin{abstract}
  We calculate the rates of formation and detection of ultracold
  Cs$_2$ molecules obtained from the photoassociation of ultracold
  atoms through the double-well $\Ogm(6\Soh + 6\Pth)$ state.  We
  concentrate on two features previously observed experimentally and
  attributed to tunneling between the two wells [Vatasescu et al 2000
  Phys. Rev. A \textbf{61} 044701].  We show that the molecules
  obtained are in strongly bound levels ($v''=5,6$) of the metastable $\aSup(6\Soh
  + 6\Soh)$ ground electronic state.
\end{abstract}

\submitto{\JPB}

\maketitle

\section{Introduction}

The control of elementary interactions between atoms or molecules in
the gas phase is a long term concern for researchers, namely in order
to find ways towards the full control of a chemical reaction, giving
the ability to choose the reaction path unambiguously from a 
well-defined internal state towards  a
desired final state. In this context, laser cooling and trapping of atoms
has opened an entirely new field of investigation, as they
can be brought almost to rest, in a well-defined internal state.
A spectacular example is provided by
the observation of Bose-Einstein condensation in alkali gases
\cite{anderson1995,bradley1995,davis1995}, recently followed by the
demonstration of quantum degeneracy in a fermionic alkali gas
\cite{demarco1999}.

Slowing and cooling of molecules, despite the inherent difficulty caused by the
complex internal molecular structure, represents an increasingly
active research field with many achievements since the first
observation of ultracold Cs$_2$ molecules by photoassociation of
laser-cooled cesium atoms \cite{fioretti1998}.  Two approaches have
been proven to be very efficient, namely Stark deceleration
\cite{bethlem2000a} and buffer gas cooling
\cite{weinstein1998a}. Other promising methods rely on phase space
filtering of a molecular beam \cite{rangwala2003}, on billiard-like
collisions \cite{elioff2003}, or on a gas expansion out of a rotating
nozzle \cite{gupta1999}. A major breakthrough has been the observation
of molecular Bose-Einstein condensates, using magnetic field
tunability of Feshbach resonances \cite{jochim2003,zwierlein2003}. Also,
two groups recently measured for the first time rates for collisions between
ultracold cesium atoms and cesium molecules \cite{staanum2006,zahzam2006}.

Among the various approaches, the photoassociation (PA) process is
very attractive, as it starts from a pair of ultracold atoms which
absorbs a photon to form an ultracold molecule in a short-lived
excited electronic state \cite{thorsheim1987}. The main drawback is
that the stabilization of the excited molecule into stable electronic
states hardly provides an ensemble of molecules in a well-defined
internal state, as it relies on spontaneous emission which populates
many vibrational levels. Some attempts to use stimulated emission have
been reported \cite{tsai1997,tolra2001}, but they were limited to the
transition into high-lying levels. Indeed, due to the poor spatial
overlap of the wave function of the photoassociated level ---
predominant at large interatomic distances --- with the wave function
of the lowest vibrational levels of the stabilized molecules, it is
very difficult to create a significant proportion of ultracold
molecules in their absolute ground state. A first indication of the
possibility to form ultracold molecules in their absolute ground state
has been provided in~\cite{nikolov2000} for K$_2$ molecules using a
two-step photoassociation process. Very recently, a multistep
excitation/deexcitation scheme has been set up to produce a fair
amount of RbCs ultracold molecules in the $v=0$ level of the
electronic ground state \cite{sage2005}.

The purpose of this paper is to describe a possible way to fill these
objectives, in the particular case of the formation of ultracold
Cs$_2$ molecules, using a single-step excitation scheme.  It results
from the cooperative action of photoassociation at large distances,
tunneling and resonant spin-orbit coupling at short distances, and
final stabilization by spontaneous decay towards a few low-lying
vibrational levels of a stable molecular state.  We first invoked such
a process for the interpretation of so-called ``giant lines'' observed
in photoassociation spectra in~\cite{vatasescu2000}, hereafter referred
to as paper~I.  One of us also performed a time-dependent analysis of
the tunneling process in such ultracold conditions, helpful for the
determination of the characteristic times of the tunneling motion
\cite{vatasescu2002}. From the present calculations involving the
available extensive photoassociative spectroscopy of Cs$_2$, we
demonstrate that the ultracold molecules are created with a very
narrow distribution of vibrational levels, peaking at $v''=5,6$ of the
theoretical potential curve of the Cs$_2$ metastable triplet
state. This could be the first example of the formation of ultracold
molecules left mainly in a single deep vibrational level via
single-photon PA.  The possibility to use this mechanism to create the initial
state of cesium dimers in ultracold collisions with cesium atoms has been
discussed in ref.\cite{staanum2006}.

We present our calculation of the photoassociation
rate, of the cold molecule formation rate, and of the molecular ion
signal resulting from the photoionization of the stabilized cold
molecules. This emphasizes the role of the detection process in the
interpretation of the strong intensity of these giant lines observed
in the photoassociation spectrum.

We start in section~\ref{sec:facts} by recalling the main facts
concerning the tunneling effects in the photoassociation of Cs$_2$.
We then present the model used to calculate the molecule formation
rates (section~\ref{sec:model}), followed by the results of the
simulations for the Cs$_2^+$ ionization signals
(section~\ref{sec:results}). Finally, concluding remarks are given in
section~\ref{sec:conclu}.

\section{Experimental and theoretical facts on giant lines and
  tunneling in the PA of cesium}
\label{sec:facts}

\begin{figure}
\center
\includegraphics[width=0.7\columnwidth]{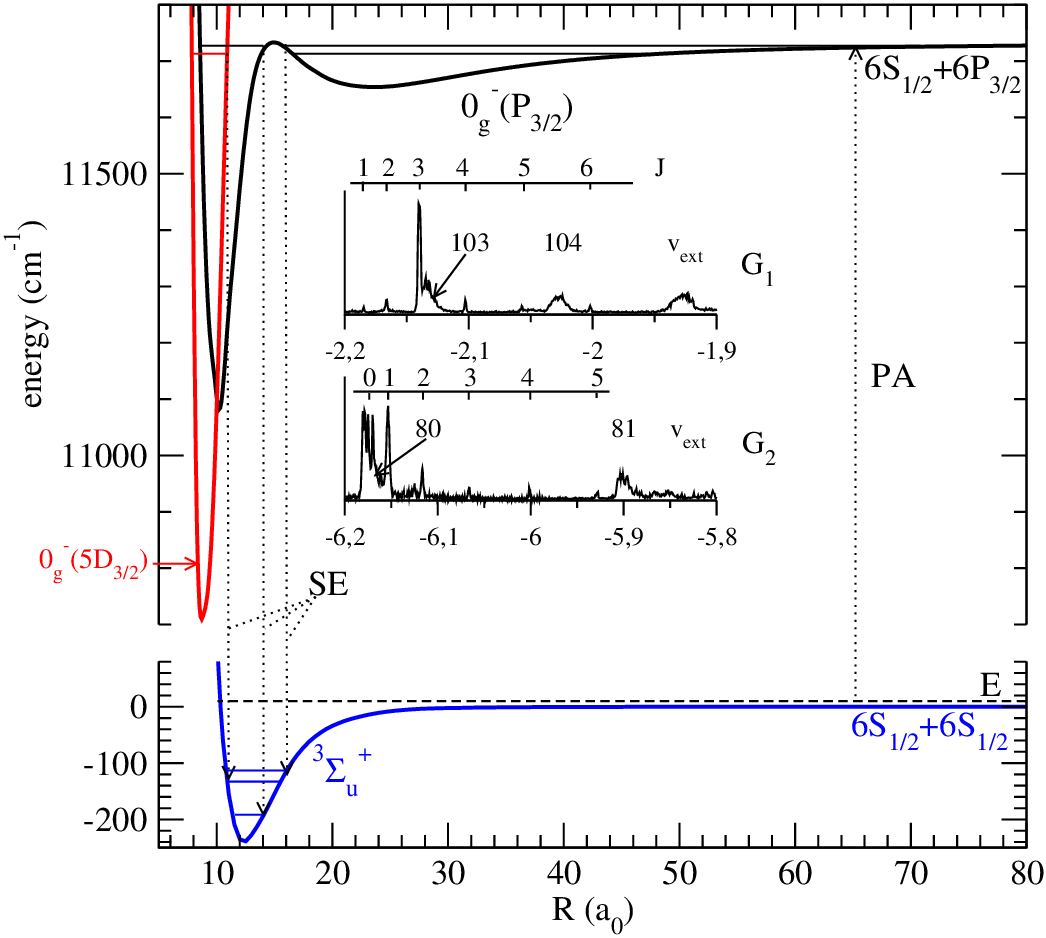}
  \caption{Scheme of the photoassociation (PA) process and subsequent
    spontaneous emission (SE) of the $\Ogm(\Pth)$ excited molecule
    to the metastable $\aSup(6\Soh+6\Soh)$ state. The
    $\Ogm(\Pth)$ state is coupled to the $\Ogm(5\Dth)$ via spin-orbit
    interaction. Vibrational levels involved in the tunneling through
    the $\Ogm(\Pth)$ barrier, as well as the deep $\aSup$ vibrational
    levels which can be reached by spontaneous decay are pictured
    with lines (not to scale).  The recorded experimental spectrum is
    recalled in the two insets, for both the G$_1$ and G$_2$ features,
    showing their large rotational structure, and the neighboring
    levels $v_{\mathrm{ext}}$ of the $\Ogm(\Pth)$ external well.}
\label{fig:potS0g}
\end{figure}

We briefly recall below the main results of paper~I.\@ The PA process
between two ultracold cesium atoms in a magneto-optical trap (MOT),
shown in figure~\ref{fig:potS0g}, is written as
\begin{eqnarray}
  \mathrm{Cs}(6\mathrm{s}^{2}\mathrm{S}_{1/2}, F=4) +
  \mathrm{Cs}(6\mathrm{s}^{2}\mathrm{S}_{1/2}, F=4) +
  \hbar(\omega_{\mathrm{D2}} - \DL) \nonumber \\ 
  \rightarrow
  \mathrm{Cs}_{2}(\Ogm(6\mathrm{s}^{2}\mathrm{S}_{1/2} +
  6\mathrm{p}^{2}\mathrm{P}_{3/2}; v,J)).
\end{eqnarray}
The pair of cold cesium atoms absorbs a photon detuned by $\DL$ to the
red of the D2 atomic transition frequency $\omega_{\mathrm{D2}}$, to
populate rovibrational levels of the external well of the
double-well-shaped $\Ogm$ molecular state correlated to the $6\Soh +
6\Pth$ dissociation limit (hereafter labeled $\Ogm(\Pth)$).  These
levels are detected by two well-established methods: first through the
fluorescence variations of the atomic trap induced by their
spontaneous decay towards bound levels or towards the dissociation
continuum of the lowest $\aSup(6\Soh+6\Soh)$ state, second through the
resonant two-photon ionization (R2PI) of the ultracold molecules
formed in this metastable triplet state into Cs$_2^+$ molecular
ions. The latter method delivers a very neat spectrum with excellent
signal-to-noise ratio. Typical experimental conditions are
temperatures between 20 and $200\ \mu\mathrm{K}$, a mean atomic
density of about $4 \times 10^{10}\ \mathrm{cm}^{-3}$ with a peak
density of $10^{11}\ \mathrm{cm}^{-3}$, the number of atoms being
estimated between 2 to $5\times 10^{7}$ atoms, and intensities of 50
up to 500~W/cm$^2$ for the the photoassociation
laser~\cite{drag2000}. The full R2PI spectrum for the $\Ogm(\Pth)$ has
been analyzed in~\cite{comparat1999}, yielding a very precise
determination of the corresponding potential curve through the
Rydberg-Klein-Rees (RKR) analysis, and through an approach based on
the asymptotic modeling of the atom-atom interactions
\cite{gutterres2002}. Vibrational levels from $v_{\mathrm{ext}}=0$ to
$v_{\mathrm{ext}}=132$ have been identified \cite{fioretti1999} in the
$\Ogm(\Pth)$ outer well, with a rotational structure (up to $J=4$)
clearly visible up to $v_{\mathrm{ext}}=72$.

In addition, two intense structures with a large rotational splitting
(hereafter referred to as ``giant lines'', following paper~I, and
labeled G$_1$ and G$_2$) are superimposed on lines associated with
levels of the $\Ogm$ external well (figure \ref{fig:potS0g}). The most
intense line within the G$_1$ (G$_2$) feature was assigned to $J=3$
($J=0, 1$) at $\DL=-2.14$~cm$^{-1}$ ($\DL=-6.15$~cm$^{-1}$) superimposed
on $v_{\mathrm{ext}}=103$ ($v_{\mathrm{ext}}=80$), while weak lines
assigned up to $J=6$ were detected providing a rotational constant
$B_v^{\mathrm{G}_1} = 137\pm 4$~MHz ($B_v^{\mathrm{G}_2}=243\pm 8$~MHz).

With the help of a coupled-channel model, the G$_1$ and G$_2$ features
have been assigned in paper~I to levels which tunnel through the
potential barrier between the two wells of the $\Ogm(\Pth)$ state, as
schematized in figure~\ref{fig:potS0g}. Their large rotational
structure is induced by the vibrational motion in two coupled
vibrational levels of the $\Ogm(\Pth)$ internal well, which is coupled
to several levels in the external well (only two of them are drawn in
figure~\ref{fig:potS0g} for better clarity). The tunneling effect,
unusual for a heavy molecule like Cs$_2$, is then an efficient
mechanism to transfer the vibrational motion of the photoassociated
state from large interatomic distances towards the inner zone: in
contrast with the long-range molecular states usually reached by PA,
spontaneous emission of the tunneling levels can stabilize the
photoassociated molecules into low vibrational levels of the $\aSup$
metastable state, creating then ultracold molecules with a rather cold
vibrational motion. Let us note that the internal state of the
molecules created has not been probed experimentally yet.

The $\Ogm(\Pth)$ outer well is represented by the asymptotic model
of~\cite{gutterres2002}.  As there is no available spectroscopic
determination of the $\Ogm(\Pth)$ inner well, this external well is
matched to the curve computed by Spies and Meyer~\cite{spies1989},
through a potential barrier whose height and position have been
adjusted in paper~I in order to reproduce the experimental
observations. The potential barrier then culminates at 2~cm$^{-1}$
above the $6\Soh+6\Pth$ dissociation limit. Furthermore, according
to~\cite{spies1989}, the $\Ogm(\Pth)$ state is coupled in the inner
well region to the next $\Ogm(6\Soh+5\Dth)$ state (hereafter labeled
$\Ogm(5\Dth)$) through a non-adiabatic coupling generated by
spin-orbit interaction. The $\Ogm(\Pth)$ and $\Ogm(5\Dth)$ curves then
exhibit a well-localized avoided crossing around $10 a_0$, which we
transformed into a real crossing by linearizing the curves around the
crossing to determine the corresponding standard Landau-Zener coupling
parameters. We imposed a finite range for the coupling using the
Gaussian form $A \exp\left[ -(r-r_0)^2/w^2 \right]$ with $r_0=10.02$,
$w=2$, and $A=0.000258$ (all in atomic units). We recall that the
absence of resonant coupling between levels of the internal wells
would result to a single tunneling level, instead of the two levels
assigned in the experiment~\cite{vatasescu2000}.

\begin{figure}
\center
\includegraphics[width=0.7\columnwidth]{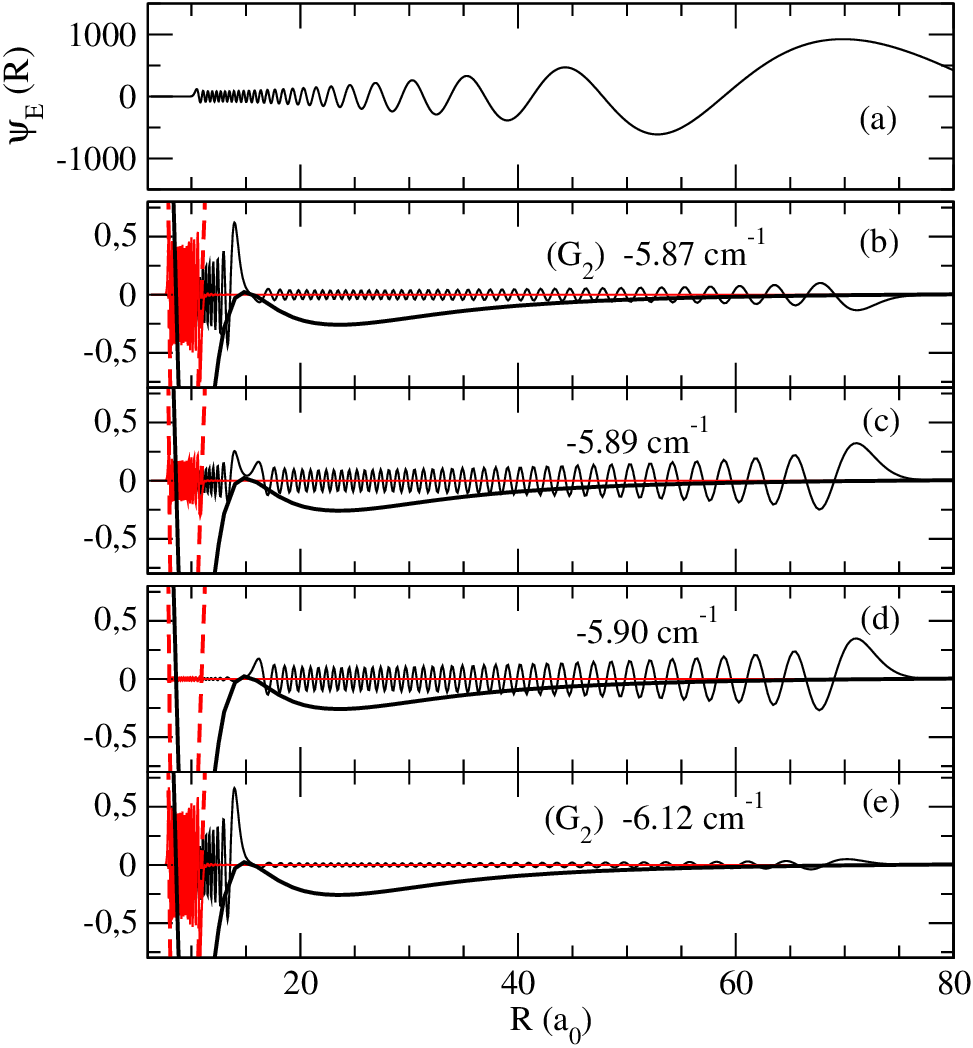}
  \caption{Two-channel vibrational wave functions
    $\chi_{\bar{v},\bar{J}}(R)$ (panels b-e) associated with the
    radial motion in the $\Ogm(\Pth)$ double well (in black) and in
    the $\Ogm(5\Dth)$ well (in red). The corresponding potentials are
    drawn with full black and dashed red lines, respectively, to make
    more visible the extension of the wave functions within the
    corresponding potentials. The $\bar{J}=5$ value (panels b and c)
    allows maximal tunneling for the G$_2$ resonance at
    $-5.87$~cm$^{-1}$, almost degenerate with a level of the external
    well at $-5.89$~cm$^{-1}$: the wave functions have significant
    amplitude over the barrier range.  The $\bar{J}=0$ case (panels d
    and e) is also shown for comparison: the energy matching is less
    favorable and the wave functions hardly penetrate the potential
    barrier. Panel (a) displays a typical radial wave function of the
    initial collisional state for a temperature of $150\
    \mu\mathrm{K}$, in the $\aSup(6\Soh+6\Soh)$ state, designed here
    with a scattering length of about $2370 a_0$.}
\label{fig:wfG2J05}
\end{figure}

For each value of $J$, the radial Schr\"{o}dinger equation is solved with
the Mapped Fourier Grid Hamiltonian (MFGH)
method~\cite{kokoouline1999}, providing accurate vibrational energies
$E_{vJ}$, wave functions $\chi_{vJ}(R)$, and rotational constants
$B_{vJ} = \langle \chi_{vJ} | \hbar^2 /(2 \mu R^2) | \chi_{vJ}\rangle$
(where $\mu=121135.83\ \mathrm{a.u.}$ is the Cs$_2$ reduced mass) of
the two coupled state.  It is well known that tunneling through a
potential barrier is very sensitive to its shape and to the position
in energy with respect to the top of the barrier. These effects are
magnified here as the two tunneling levels of the $\Ogm(\Pth)$ and
$\Ogm(5\Dth)$ internal wells, hereafter referred to as
$v_{6\mathrm{P}}$ and $v_{5\mathrm{D}}$, respectively, interact
resonantly with the dense --- but not continuous --- energy level
spectrum of the external well. Our calculations predict that the tunneling effect
will be observable in the detection signal if the tunneling probability is 
maximum, which is achieved by
adjusting the relative position of the two internal wells to create an
almost half-and-half mixing of the $v_{6\mathrm{P}}$ and
$v_{5\mathrm{D}}$ radial wave functions. Our model actually presents
many tunable parameters for such an adjustment: indeed, the internal
well of the potentials, as well as their coupling, are unknown
experimentally, so that their fine tuning is quite tedious. We also
noted that the tunneling effect is very sensitive to the value of
$J$. In order to facilitate the convergence, we stopped the
adjustment when we found maximal tunneling for an arbitrary $J$
value.  In other words, we also considered the rotational number $J$
as a tunable parameter, which we will label as $\bar{J}$ in the
following. An example of the resulting wave functions is shown in
figure~\ref{fig:wfG2J05}), where the maximal tunneling effect is found
for the value $\bar{J}=5$. This is clearly not in agreement with the
rotational quantum number of the experiment, but we only need
\emph{radial} wave functions for the rate calculations of the next
section, which will be evaluated without any $J$ dependence. The
corresponding numerical data is available on request from the authors.

\begin{figure}
\center
\includegraphics[width=0.7\columnwidth]{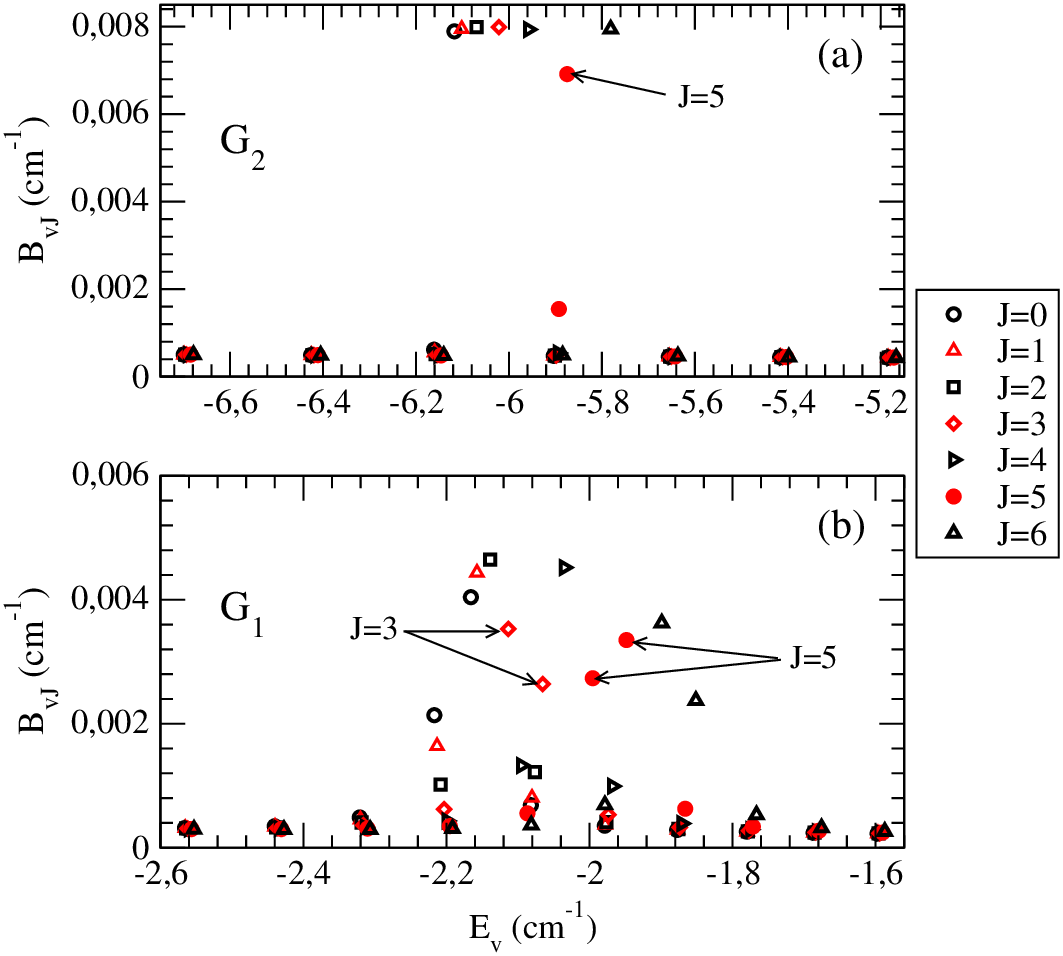}
  \caption{Rotational constants $B_{vJ}=\langle \chi_{\bar{v},\bar{J}}
    |\hbar^2(2 \mu R^2)| \chi_{\bar{v},\bar{J}}\rangle$ for
    rovibrational levels of the coupled $\Ogm(6\Soh+6\Pth)$ and
    $\Ogm(6\Soh+5\Dth)$ potentials, as function of their binding
    energy. (a) G$_2$ structure; (b) G$_1$ structure. The levels with
    an important tunneling effect are marked by arrows.}
\label{bvJ.eps}
\end{figure}

Even if in the adjustment procedure above we used $\bar{J}$ as a
parameter, figure~\ref{bvJ.eps} shows that the selectivity of the
tunneling process is very strongly dependent on the rotational
level. The value of $B_{vJ}$ defined above is very sensitive to the
repartition of the wave function inside the two wells, and panel (a)
in figure~\ref{bvJ.eps} reflects the pattern shown in
figure~\ref{fig:wfG2J05}, where the value for $\bar{J}=5$ is
intermediate between the rotational constant of levels in the internal
well ($\approx 0.008$~cm$^{-1}$) and of levels of the external well
($\approx0.0005$~cm$^{-1}$). Similarly, the $\bar{J}=3$ line for G$_1$
also reflects the extension of the wave function over both
wells. The $\bar{J}=5$ line displays the same behavior, which can be
understood since the rotational structure of G$_1$ has the same magnitude
as the energy spacing between vibrational levels of the external well,
so that it is resonant with $v_{\mathrm{ext}}=104$ in our
calculations.  This aspect was not present in the model we set up in
paper~I and confirms what is observed experimentally. Indeed, due to
the ultracold temperature, the s-wave regime is expected to dominate
the initial collision between the two Cs atoms (yielding $J=0, 1, 2$)
with probably a small contribution of the p wave (yielding $J=1, 2,
3$).  The presence of higher values of $J$ at such low temperatures is
not yet fully understood~\cite{fioretti1999}.  The relative
intensities of the $J=0,1,2$ experimental lines approximately reflects
their $2J+1$ degeneracy, and the $J=2$ line has indeed the strongest
intensity for most of the PA lines assigned to the levels of the
$\Ogm(\Pth)$ external well. The lines with larger $J$ values are
weaker, due to the small contribution of higher partial waves involved
in the collision. We see that this hierarchy is not preserved for
G$_1$ and G$_2$: the intense lines are associated with those rotational
levels which are indeed resonant with a level of the internal well,
i.e., $J=3$ and $J=0,1$, respectively.

\section{Photoassociation and cold molecule formation rates for
  tunneling levels}
\label{sec:model}

The metastable $\aSup(6\Soh+6\Soh)$ state is represented by the
potential curve of~\cite{foucrault1992}, matched beyond $20a_0$ to the
standard asymptotic expansion $\sum{C_n/R^n}$ ($n=6,8,10$), with $C_n$
coefficients taken from~\cite{marinescu1994}. The repulsive wall of
the potential is slightly changed in order to reproduce a large and
positive scattering length~\cite{kerman2001}. We work here with
$a_{\mathrm{T}} = 2370 a_0$, which satisfactorily reproduces the
intensity envelope of the PA spectrum of~\cite{fioretti1999}. The
initial scattering radial wave function is computed through a standard
Numerov integration at an energy $E/k_{\mathrm{B}}T = 150\
\mu\mathrm{K}$ above the $6\Soh+6\Soh$ asymptote, ignoring the
hyperfine structure (figure~\ref{fig:wfG2J05}a). Let us note that the
accuracy of the $a_{\mathrm{T}}$ and $C_n$ values is not crucial for
the following calculation of the rates, which can be determined
experimentally typically within a factor of 2.

The photoassociation and cold molecule formation rates are evaluated
according to the perturbative model already presented in previous
papers~\cite{drag2000,pillet1997,dion2002}.  Briefly, the
photoassociation rate $R_{\mathrm{PA}}(\bar{v},\bar{J};T)$ per atom
(expressed in s$^{-1}$) for an initial continuum state with energy
$E=k_{\mathrm{B}}T$ and wave function $\psi_E(R)$ into a vibrational
level $\bar{v}$ with a radial wave function
$\chi_{\bar{v},\bar{J}}(\Ogm; R)$ of the coupled $\Ogm$ states (where
$\bar{v}$ stands for the quantum number $v$ of the $\Ogm(\Pth)$
external well levels, or for G$_1$ or G$_2$), at a detuning $\Delta_v$
below $6\Soh+6\Pth$ limit is expressed as
\begin{equation}
  R_{\mathrm{PA}}(\bar{v},\bar{J};T) = \left(\frac{3}{2\pi}
  \right)^{3/2}\frac{h}{2} n_{\mathrm{at}} \lambda^3_{\mathrm{th}}
  \mathrm{e}^{-\frac{E}{k_{\mathrm{B}} T}} \mathcal{A} 
  K^2  \left| \left\langle \psi_E  \left|
    \chi_{\bar{v},\bar{J}}(\Ogm) \right\rangle \right. \right|^2,
\label{eq:parate}
\end{equation}
where $n_{\mathrm{at}}$ is the atomic density and
$\lambda_{\mathrm{th}} = h\sqrt{1/(3\mu k_B T)}$ is the thermal de
Broglie wavelength. The atomic Rabi frequency 2$K$ is related to the
intensity $I$ of the laser through $K^2 = \frac{\Gamma}{8}
\frac{I}{I_0}$, where $\Gamma/2\pi=5.22$~MHz is the natural width of
the $6\Pth$ atomic level. At the PA wavelength considered,
$\lambda_{\mathrm{PA}}$, the saturation intensity $I_0=\pi hc
\Gamma/(3\lambda_{\mathrm{PA}}^3)$ is $1.1\ \mathrm{mW/cm}^{2}$.  The
angular factor $\mathcal A=125/3888$ includes hyperfine degeneracies,
assuming an initial $(F=4)+(F=4)$ hyperfine state. As mentioned
above, we actually solve the Schr\"{o}dinger equation for every chosen
value of the rotational quantum number $\bar{J}$ taken as an effective
parameter, so we don't include the degeneracy factor $2J+1$ in the
rate formulas.

\begin{figure}
\center
\includegraphics[width=0.7\columnwidth]{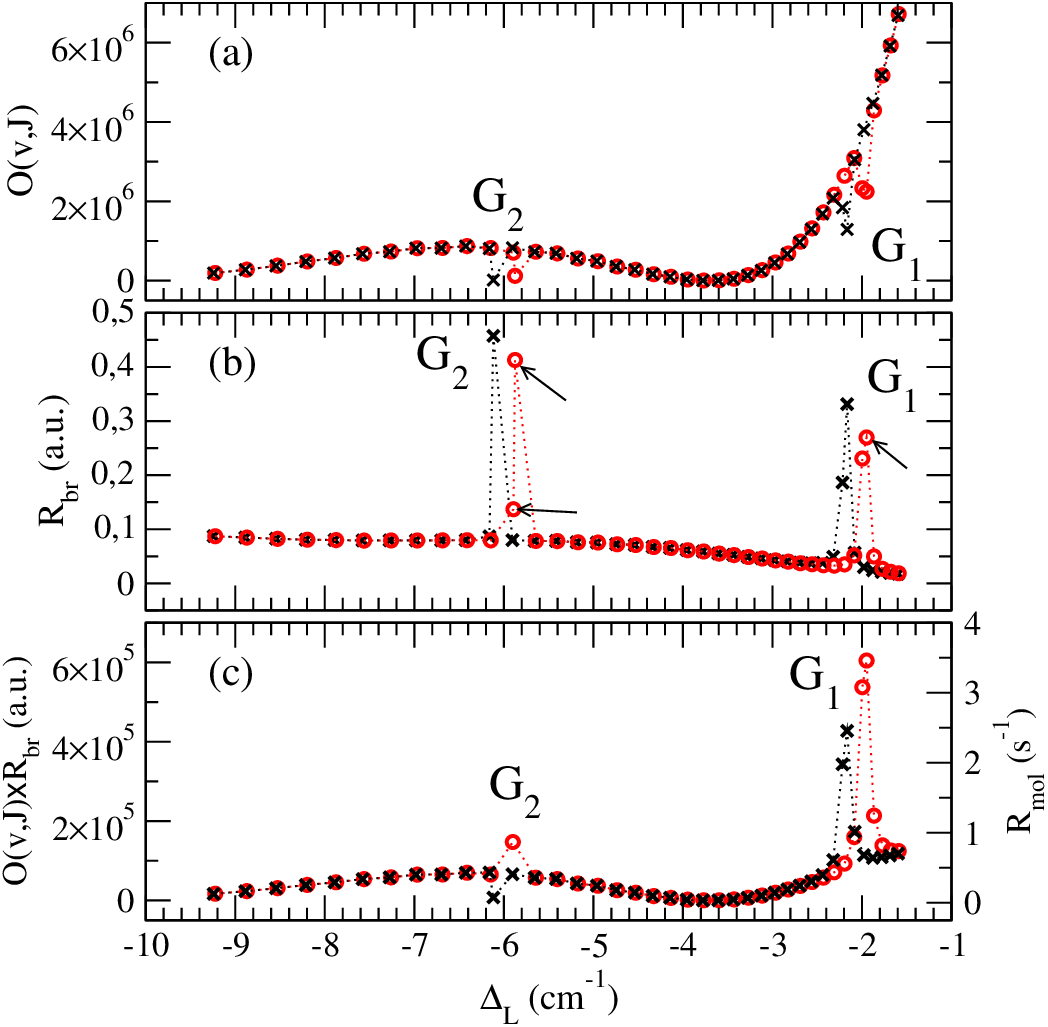}
  \caption{(a) Overlap $O(E,\bar{v},\bar{J})$ between the continuum
    wave function and vibrational wave functions of coupled $\Ogm$
    potentials lying in the detuning interval between $-9$ and $-1.5\
    \mathrm{cm}^{-1}$. (b) Branching ratio
    $R_{\mathrm{br}}(\bar{v},\bar{J})$ of the same levels to the
    $\aSup$ vibrational states. (c) Product of the two previous
    quantities (left axis) and corresponding cold molecule formation
    rate $R_{\mathrm{mol}}(\DL)$ (right axis) assuming an atomic
    density $n_{\mathrm{at}}=10^{11}$~cm$^{-3}$ and a PA laser
    intensity $I=100\ \mathrm{W/cm}^2$. Note that in this panel the
    contribution of the nearby levels at $-5.87$~cm$^{-1}$ and
    $-5.89$~cm$^{-1}$ shown in figure~\ref{fig:wfG2J05} are added
    together into a single rate value around $-5.9$~cm$^{-1}$.  Both
    $\bar{J}=0$ (black crosses) and $\bar{J}=5$ (red open circles) are
    shown. The levels corresponding to G$_1$ and G$_2$ are
    indicated. The arrows identify the levels for which we calculated
    the vibrational distribution of the produced cold molecules
    displayed in the figure~\ref{fig:G1-2J5dist}.}
\label{fig:ov-J05}
\end{figure}

The two components of the PA level wave function $\left|
  \chi_{\bar{v},\bar{J}}(\Ogm;R) \right\rangle$ (see
figure~\ref{fig:wfG2J05}) are denoted by
$\chi_{\bar{v},\bar{J}}^{6\mathrm{P}}(R)$ and
$\chi_{\bar{v},\bar{J}}^{5\mathrm{D}}(R)$, such that
\begin{equation}
  O(E,\bar{v},\bar{J}) \equiv \left| \left\langle \psi_E \left|
        \chi_{\bar{v},\bar{J}}(\Ogm) \right\rangle \right. \right|^2 =
  \left| \left\langle \psi_E \left| \chi_{\bar{v},\bar{J}}^{6\mathrm{P}}
      \right\rangle \right. + \left\langle \psi_E \left|
        \chi_{\bar{v},\bar{J}}^{5\mathrm{D}} \right\rangle \right. \right|^2 .
\label{eq:paov}
\end{equation}
Following~\cite{dion2001}, the rate $R_{\mathrm{mol}}(\DL)$ for cold
molecule formation after PA in the $\bar{v}$ level is obtained by
multiplying the PA rate with the branching ratio
$R_{\mathrm{br}}(\bar{v},\bar{J})$ of the $\bar{v}$ level towards the
bound levels $v''(\aSup)$) of the metastable triplet state, and
neglecting here the $R$ dependence of the dipole transition function
for the spontaneous decay step,
\begin{equation}
R_{\mathrm{mol}}(\DL)= R_{\mathrm{PA}}(\DL) R_{\mathrm{br}}(\bar{v},\bar{J}),
\label{eq:molrate}
\end{equation}
with
\begin{equation}
R_{\mathrm{br}}(\bar{v},\bar{J}) = \sum_{v''}  \left| \left\langle
    \chi_{\bar{v},\bar{J}}(\Ogm) \left| \phi_{v''}(\aSup)
    \right\rangle \right. \right|^2.
\label{eq:brrate}
\end{equation}
Figure~\ref{fig:ov-J05}a displays the overlap between the continuum
wave function and the bound states of coupled potentials,
equation~(\ref{eq:paov}), lying in the energy interval between $-9$
and $-1.5$~cm$^{-1}$, for both $J=0$ and $J=5$. As expected, G$_1$ and
G$_2$ have smaller overlap with the initial continuum than the
external well levels, due to the weaker probability of wave functions
to be localized at large distances. In contrast with the PA rate, the
branching ratio of the G$_1$ and G$_2$ wave functions is enhanced due
to their good localization at short distances
(figure~\ref{fig:ov-J05}b).  As it was often emphasized, a good
production of cold molecules requires a favorable ratio between
free-bound and bound-bound transitions. The product
$O(E,\bar{v},\bar{J}) \times R_{\mathrm{br}}(\bar{v},\bar{J})$ is
represented in figure~\ref{fig:ov-J05}c. If for G$_1$ the efficiency
seems certain, we see that for G$_2$ this balance is fragile and
depends indeed on the $\bar{J}$ value which makes tunneling
effective. On the right axis of this figure is reported the rate for
cold molecule formation, equation~(\ref{eq:molrate}), for typical
experimental conditions $n_{\mathrm{at}}=10^{11}$~cm$^{-3}$ and a PA
laser intensity $I= 100\ \mathrm{W/cm}^2$. It is found two times and
eight times larger than the rate for the $v_{\mathrm{ext}}$ levels for
G$_2$ and G$_1$, respectively.  Even if in the experiment the
intensities of the G$_2$ and G$_1$ lines are found to be comparable
(see insets in figure~\ref{fig:potS0g}), our results confirm that the
cold molecule formation rate is indeed larger than the rate for the
$v_{\mathrm{ext}}$ levels. The relative intensity of the G$_2$ and
G$_1$ lines may be influenced by the two-photon ionization used for
the detection (see next section). Let us note also that the rate
computed for the $v_{\mathrm{ext}}$ levels is in agreement with the
one reported in~\cite{drag2000}, where $R_{\mathrm{mol}} \approx 0.2\
\mathrm{s}^{-1}$ was measured at a detuning of 7~cm$^{-1}$, for a PA
laser intensity $I= 55\ \mathrm{W/cm}^2$.

\begin{figure}
\center
\includegraphics[width=0.7\columnwidth]{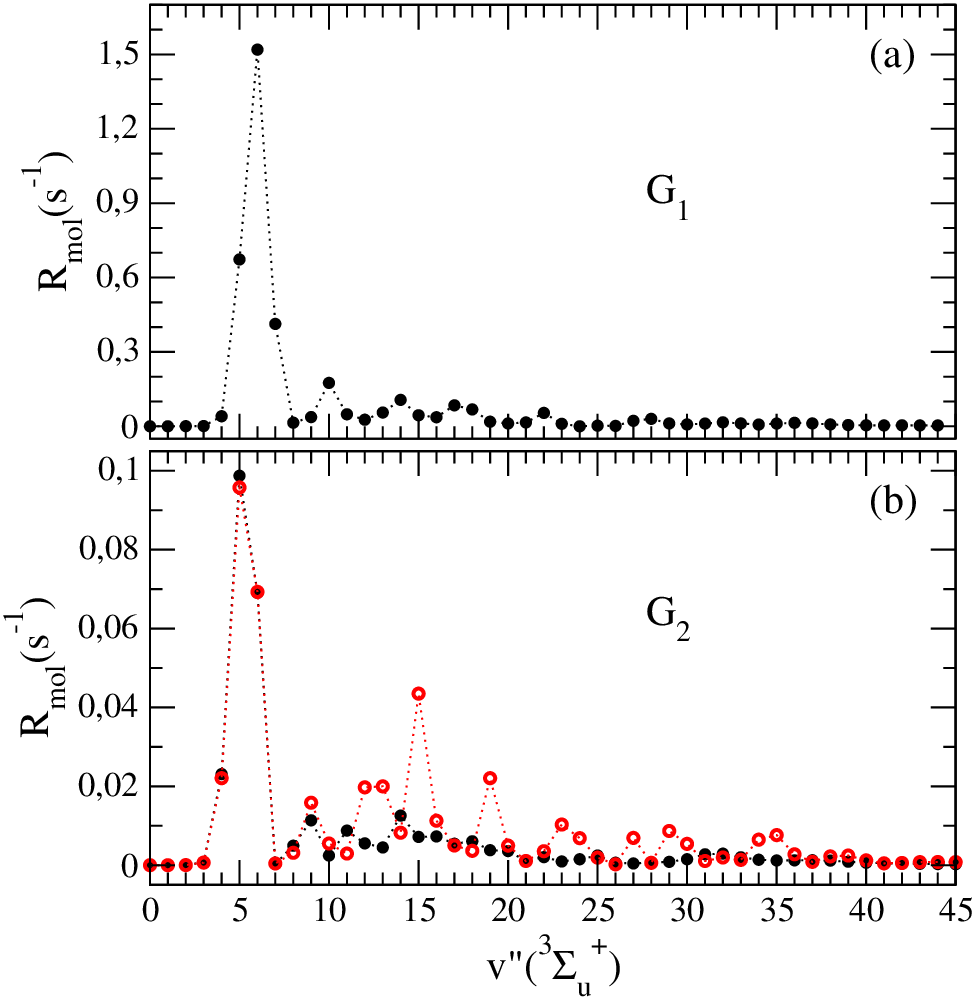}
  \caption{Rate of formation of cold molecules in the vibrational
    levels $v''(\aSup)$, via photoassociation into (a) the G$_1$ and
    (b) the G$_2$ states. In the latter case, rates for both levels
    contributing to G$_2$ are given, i.e., the level at
    $-5.87$~cm$^{-1}$ tunneling from the inner wells (black closed
    circles) and the external level tunneling from the outer well at
    $-5.89$~cm$^{-1}$ (red open circles). The calculation is made for
    $n_{\mathrm{at}}=10^{11}$~cm$^{-3}$ and an intensity $I= 100\
    \mathrm{W/cm}^2$ for the PA laser.}
\label{fig:G1-2J5dist}
\end{figure}

The rates for formation of cold molecules in individual levels $v''$
of the metastable $\aSup$ state,
\begin{equation}
  R_{\mathrm{mol}}(\DL,v'') = R_{\mathrm{PA}}(\DL) \left| \left\langle
      \chi_{\bar{v},\bar{J}}(\Ogm) \left| \phi_{v''}(\aSup)
      \right\rangle \right. \right|^2,
\label{eq:molrate_v}
\end{equation}
are represented in figure~\ref{fig:G1-2J5dist} for the same range of
detunings and same experimental conditions as in
figure~\ref{fig:ov-J05}c, for both the G$_1$ and G$_2$ features,
considering the levels indicated by arrows in figure~\ref{fig:ov-J05}b
corresponding to $\bar{J}=5$. Both vibrational distributions peak
markedly at $v''=5,6$, confirming the efficiency of the PA into the
tunneling resonances to produce cold molecules in very low vibrational
levels.

\section{Modeling the ionization signal}
\label{sec:results}

In cesium PA experiments~\cite{fioretti1998,vatasescu2000}, the cold
molecules are detected via a two-photon ionization process, where the
first photon is resonant with rovibrational levels of molecular states
correlated to the $6\Soh+5\Dth$ dissociation limit. Therefore, the
ionization signal is very sensitive to the wavelength chosen for the
ionization laser. Following our previous studies of this
process~\cite{dion2002,dion2001}, we simulate the contributions of
G$_1$ and G$_2$ to the Cs$_2^+$ ions signal obtained after PA in a
$\Ogm(6\Soh+6\Pth)$ level detuned by $\DL$, spontaneous emission
towards $v''(\aSup)$ bound levels, and ionization with a photon of
frequency $\nu_{\mathrm{ion}}$ of these levels through absorption into
levels $v'$ of the $(2)^3\Pi_{\mathrm{g}}(6\Soh+5\mathrm{D}_{5/2})$
potential. In this model, the second step of the ionization process is
described as a uniform ionization probability of the $v'$ levels into
Cs$_2^+$ ions, and will not be considered here. The ionization signal
$S_{\mathrm{ion}}(\DL)$ is expressed as
\begin{equation}
S_{\mathrm{ion}}(\DL) = N_{\mathrm{PA}}(\DL) \sum_{v''}
P_{\mathrm{ion}}(v'') \left| \left\langle \chi_{\bar{v},\bar{J}}(\Ogm)
    \left| \phi_{v''}(\aSup) \right\rangle \right. \right|^2.
\label{eq:sions}
\end{equation}
As in the cold molecules rate calculation, we neglect the $R$
dependence of the dipole transition moment involved in the spontaneous
decay step. In equation~(\ref{eq:sions}), the number of molecules
$N_{\mathrm{PA}}(\DL)$ accumulated during the photoassociation step is
defined by multiplying the PA rate $R_{\mathrm{PA}}(\DL)$ with the
number of atoms $n_{\mathrm{PA}}$ in the photoassociation area with
the residence time $t_{\mathrm{PA}}$ of the cold molecules within the
trapping region:
\begin{equation}
  N_{\mathrm{PA}}(\DL)=R_{\mathrm{PA}}(\DL) n_{\mathrm{PA}} t_{\mathrm{PA}}.
\end{equation}
In the Cesium PA experiment, typical values for these parameters are
$n_{\mathrm{PA}}=5 \times 10^{7}$ and $t_{\mathrm{PA}} \approx 10$~ms.
The excitation probability $P_{\mathrm{ion}}(v'')$ of the $v''$ levels
into vibrational levels $v'$ of the
$(2)^3\Pi_{\mathrm{g}}(6\Soh+5\mathrm{D}_{5/2})$ potential is
calculated as
\begin{equation}
  P_{\mathrm{ion}}(v'') = \sum_{v'} \left| \left\langle
      \phi_{v'}(^3\Pi_{\mathrm{g}}) \right| D \left| \phi_{v''}(\aSup)
      \right\rangle \right|^2 f(\nu_{v''v'}),
\end{equation}
where $D$ is the $R$-dependent dipole moment for the $\aSup \to 
{^3\Pi_{\mathrm{g}}}(6s+5\mathrm{D}_{5/2})$ transition calculated
in~\cite{spies1989}, and%
\begin{equation}
  f(\nu_{v''v'}) = \exp \left[ - \ln(2)
    \frac{(\nu_{v''v'}-\nu_{\mathrm{ion}})^2}{(\delta \nu)^2} \right]
\end{equation}
is a parametric function accounting for the resolution of the
experiment, estimated at $\delta \nu \approx 30$~GHz \cite{dion2002},
and $\nu_{v''v'}$ is the transition frequency between $v''$ and $v'$
levels.

\begin{figure}
\center
\includegraphics[width=0.7\columnwidth]{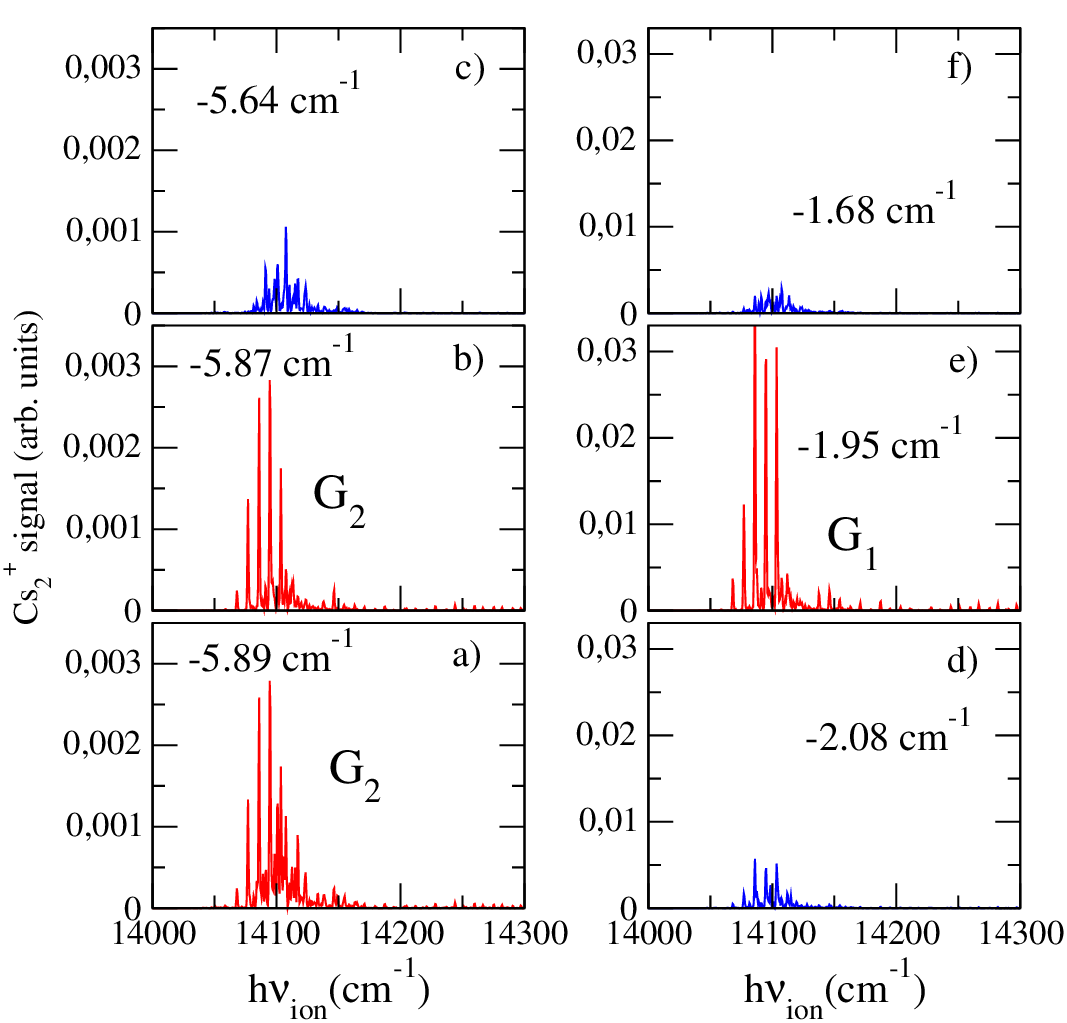}
  \caption{Cs$_2^+$ ions signal (in arbitrary units) as a function of
    the frequency of the ionization laser, calculated with
    equation~(\ref{eq:sions}) for detunings $\DL$ corresponding to
    G$_1$ (right column) and G$_2$ (left column).  Signals
    corresponding to photoassociation in vibrational levels located
    close to G$_1$ and G$_2$ and belonging to the external well of the
    $\Ogm(6\Soh+6\Pth)$ potential are also shown. The energy of the
    $6\Soh+5\mathrm{D}_{5/2}$ dissociation limit is taken as being
    $14597.08\ \mathrm{cm}^{-1}$ above the $6\Soh+6\Soh$ origin.}
\label{fig:ionJ5}
\end{figure}

\begin{figure}
\center
\includegraphics[width=0.7\columnwidth]{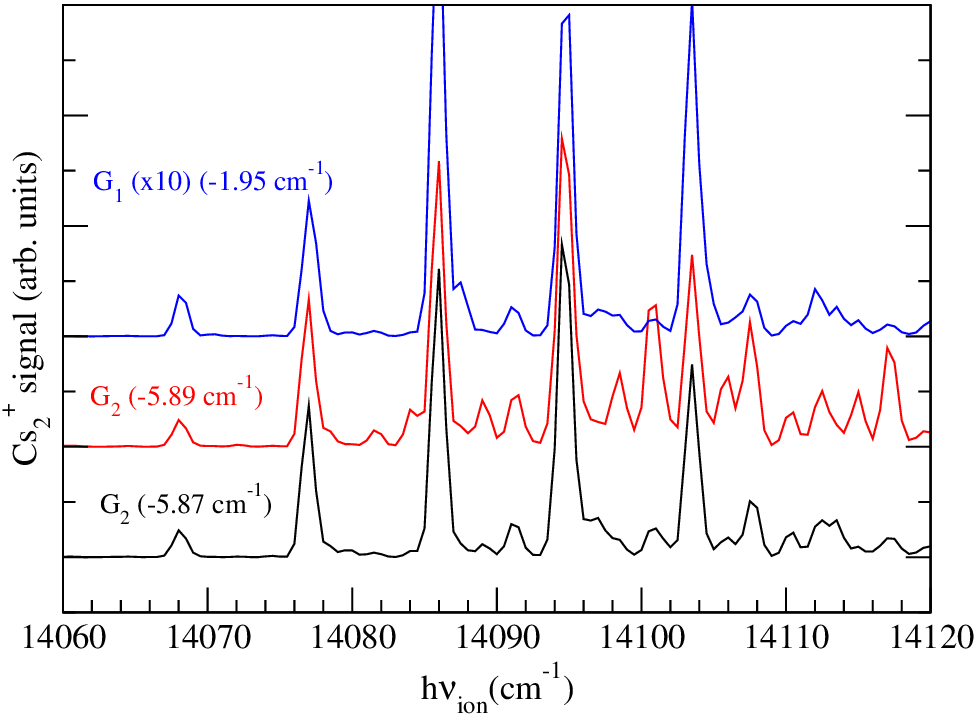}
  \caption{Same as figure~\ref{fig:ionJ5}: we superimposed predicted ion signals
    from panels a, b, and e of this figure, slightly shifted from each other for
    better visibility.  Note that the signal for
    G$_1$ is shown at 1/10th of its actual value.}
  \label{fig:ionJ5_overlay}
\end{figure}

The ion signals as a function of the ionization frequency
$\nu_{\mathrm{ion}}$ obtained for G$_1$ and G$_2$ are displayed in
figure~\ref{fig:ionJ5}, together with those corresponding to PA in
neighboring vibrational levels $v_{\mathrm{ext}}$ of the
$\Ogm(6\Soh+6\Pth)$ external well. As expected, large amounts of
molecular ions are detected only for specific frequencies
corresponding to the resonance condition in the first step.  The
signals coming from G$_1$ and G$_2$ are clearly more intense than
those of the surrounding levels, in agreement with the experimental
spectrum~\cite{vatasescu2000}, because of the increased efficiency for
forming cold molecules in bound levels of the triplet state.  The
signal corresponding to G$_1$, having a maximum of $0.03$ in
figure~\ref{fig:ionJ5}e, is five times bigger that the simulated G$_2$
signal, whose maximum reaches $0.006$ if we take the sum of the two
contributing levels (figures~\ref{fig:ionJ5}a,b). This ratio between
the intensities of the simulated ``giant lines'' is larger than in the
experimental spectrum where the ratio is only two~\cite{fioretti1999}.
This could be due to the fact that the ionization spectra for the
tunneling resonances do not match perfectly: the differences in
vibrational distributions $v''(\aSup)$ of G$_1$ and G$_2$ (see
figure~\ref{fig:G1-2J5dist}) make it possible to chose an ionization
frequency (e.g., $h \nu_{\mathrm{ion}} \approx 14101$ or
14117~cm$^{-1}$) such that the ratio between the two ion signals is
greatly modified, as shown in figure~\ref{fig:ionJ5_overlay}.

\section{Conclusion}
\label{sec:conclu}
 
As demonstrated through several
studies~\cite{fioretti1998,vatasescu2000,dion2001}, cesium atoms are
well-suited for ultracold molecule formation through photoassociation,
mainly due to the peculiarity of some of the excited states of the
Cs$_2$ molecule.  Of particular interest is the $\Ogm(\Soh+\Pth)$
state, which is composed of two potential wells separated by a barrier
whose height culminates at an energy close to the dissociation
asymptote.  This allows for tunneling between the two wells, enabling
a pair of atoms initially far apart to come close together in a
photoassociated molecule.

We have shown here that this tunneling process leads to strongly bound
ultracold cesium molecules in the metastable $\aSup(\Soh+\Soh)$ ground
electronic state.  Indeed, it is mostly vibrational levels around $v''
=5,6$ that are populated, in contrast to what can be reached for
non-tunneling states of the external $\Ogm(\Pth)$ well ($v'' > 23$
\cite{drag2000}).  The tunneling efficiency is seen to be markedly
dependent upon the rotational state $J$, in agreement with what was
observed experimentally~\cite{vatasescu2000}.

Finally, we have investigated that the experimental detection scheme,
resonant two-photon ionization, influences the observation of
ultracold Cs$_2$ molecules.  Evidence would suggest that the detection
process affects the relative intensities seen in the signal, as was
the case previously~\cite{dion2001}.  In other words, the molecules
produced are not all ionized with the same efficiency, and it is
therefore important to consider this step in the analysis of
photoassociation experiments.

\ack
This work has been performed in the framework of the Research and
Training Network of the European Union ``Cold Molecules'' (contract
number HPRN-CT-2002-00290). O.D. and M.V. acknowledge partial support from
the INTERCAN and QUDEDIS networks of the European Science Foundation.

\section*{References}


\end{document}